\documentclass{article}
\usepackage{spconf,amsmath,graphicx}
\usepackage{CJK}
\usepackage[top=25mm, bottom=25mm, left=19mm, right=19mm]{geometry}
\usepackage{algorithm}
\usepackage{algorithmicx}
\usepackage{algpseudocode}
\usepackage{amsmath}
\usepackage{multirow}
\usepackage{float}

\title{Insights into End-to-End Learning Scheme for Language Identification}
\name{Weicheng Cai $^{1}$, Zexin Cai$^{1}$ Wenbo Liu$^{3}$, Xiaoqi Wang$^{4}$, and Ming Li$^{1,2}$\sthanks{This research was funded in part by the National Natural Science Foundation of China (61401524,61773413), Natural Science Foundation of Guangzhou City (201707010363), Science and Technology Development Foundation of Guangdong Province (2017B090901045), National Key Research and Development Program (2016YFC0103905).}}

\address{$^1$School of Electronics and Information Technology, Sun Yat-sen University, Guangzhou, China\\
	$^2$Data Science Research Center, Duke Kunshan University, Kunshan, China\\
	$^3$Department of Electrical and Computer Engineering, Carnegie Mellon University, Pittsburgh, USA\\
	$^4$Jiangsu Jinling Science and Technology Group Limited\\
	{\small \tt ml442@duke.edu}}

\begin{document}

	\ninept
	\maketitle
	\begin{abstract}

	A novel interpretable end-to-end learning scheme for language identification is proposed. It is in line with the classical GMM i-vector methods both theoretically and practically.  In the end-to-end pipeline, a general encoding layer is employed on top of the front-end CNN, so that it can encode the variable-length input sequence  into an utterance level vector  automatically.  After comparing with the state-of-the-art GMM i-vector methods, we give insights into CNN, and reveal its role and effect in the whole pipeline. We further introduce  a general encoding layer, illustrating the reason why they might be appropriate for language identification. We elaborate on several typical encoding layers, including a temporal average pooling layer, a recurrent encoding layer  and a novel learnable dictionary encoding layer. We conducted experiment on NIST LRE07 closed-set task,  and the results show that our proposed end-to-end systems achieve state-of-the-art performance.

		\end{abstract}
	
	\section{Introduction}
	\label{sec:intro}

	Language identification (LID) can be defined as a utterance level paralinguistic speech attribute classification task, in compared with automatic speech recognition, which is a ``sequence-to-sequence" tagging task. There is  no constraint on the lexicon words thus the training utterances  and testing segments may have completely different contents \cite{Kinnunen2010An}. The goal, therefore, might to find a robust and time-invariant utterance level vector representation describing the distribution of local features. 
	\begin{figure*}[tb]
		\centering
		\includegraphics[width=0.93\textwidth]{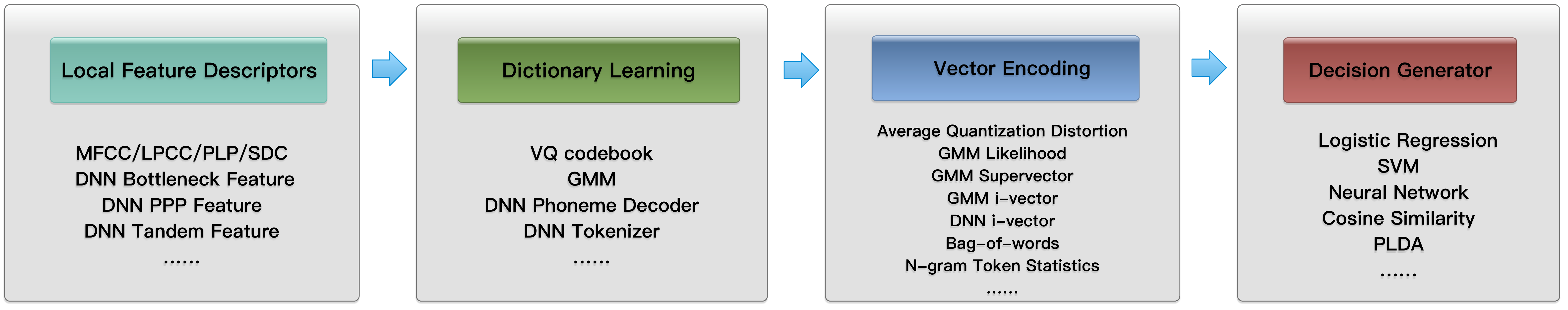}
		\caption{Four main steps in the conventional processing pipeline}\label{fig:offtheshelf}
	\end{figure*}

	In recent decades, we have witnessed that the classical Gaussian Mixture Model (GMM) i-vector approach dominates  many kinds of paralinguistic speech attribute recognition fields for its superior performance, simplicity and efficiency \cite{dehak2010front,Dehak2011Language}. As shown in Fig. \ref{fig:offtheshelf},  the processing pipeline contains four main steps as follows: 
	
	\begin{itemize}
		\setlength{\itemsep}{0pt}
		\setlength{\parsep}{0pt}
		\setlength{\parskip}{0pt}
		\item Local feature descriptors,  which manifest as a variable-length feature sequence, include hand-crafted acoustic level features, such as log mel-filterbank energies (Fbank), mel-frequency cepstral coefficients (MFCC), perceptual linear prediction (PLP), shifted delta coefficients (SDC) features \cite{Kinnunen2010An,6451097}, or automatically learned phoneme discriminant features from deep neural networks (DNN), such as bottleneck features \cite{matejka2014neural, Song2015Deep}, phoneme posterior probability (PPP) features \cite{Li2016Generalized}, and  tandem features \cite{Richardson2015Deep}. 
		
		\item Dictionary, which contains several orderless center components( or units, words), including  vector quantization (VQ) codebooks learned by K-means \cite{Soong1985Report}, a universal background model (UBM) learned by GMM \cite{Reynolds1995Robust,Reynolds2000Speaker} or a supervised phonetically-aware acoustic model learned by DNN \cite{yun_icassp14,li2014interspeech}.
		\item Vector encoding. This procedure  aggregates the  variable-length feature sequence  into an utterance level  vector representation,  based on the learned dictionary mentioned above. The typical examples are the well-known GMM Supervector \cite{campbell2006support}, the classical  GMM i-vector \cite{dehak2010front} or  recently popular DNN i-vector \cite{Snyder2016Time}.
		\item Decision generator, includes logistic regression (LR), support vector machine (SVM), and neural network for closed-set identification, cosine similarity or probabilistic linear discriminant analysis (PLDA) \cite{Prince2007Probabilistic,Kenny2010Bayesian} for open-set verification.
	\end{itemize}

	Despite the great success, the baseline GMM i-vector methods remain room to be improved: First, the front-end process to get i-vectors is totally 
	unsupervised. Thus, the i-vectors we have extracted may ignore some patterns that might be useful for the back-end classification. Second, the process to obtain i-vectors is comprised of a series hand-crafted or ad-hoc algorithmic components. Once feature is extracted or front-end model is trained, they are fixed that can't benefit from subsequent steps. For example, we often let the extracted i-vectors fixed and tune the back-end algorithm for better performance. However, no matter how superior the back-end algorithm is, we may still suffer from a local minima determined by the discrimination carried on i-vector. 
	

	Therefore, the recent progress towards end-to-end learning opens up a new area for exploration \cite{lopez2014automatic,gonzalez2014automatic,Snyder2017Deep,Geng2016End, Gelly2016A,Jin2017End,1705.02304}. The work in \cite{1705.02304} introduced deep convolutional neural Network (CNN) from image recognition, and proved its success on their own private speaker recognition database. In both \cite{Snyder2017Deep,1705.02304}, similar temporal average pooling (TAP) layer is adopted within their neural network architectures. 

	To the best of our knowledge, many of existing end-to-end works perform excellent on trial-and-error experiments, however, there is no very clear explanation about why they perform so well, or how they might be further improved. In this paper, we explore both issues.  As the first contribution, we give insights into a general learning scheme for end-to-end LID, with an emphasis on  CNN and a  encoding layer. We explore how CNN can transform the  variable-length input feature sequence  into high-level representation, acting as an automatic front-end feature extractor. Although the contextual information in convolutional receptive field is captured, the feature extracted by CNN is still with temporal order. The remaining question is: how to aggregate them together over the entire and potentially long duration length? Concerning about that, as our second contribution, we introduce a general encoding layer. We provide its general structure and elaborate on several typical vector encoding layers, including a simple TAP layer, a recurrent encoding layer, and a learnable dictionary encoding (LDE) layer as well.

	\section{Insights}
	\label{sec:insigths}

	\subsection{End-to-end learning scheme}

\begin{figure}[tb]
		\centering
		\includegraphics[width=\columnwidth]{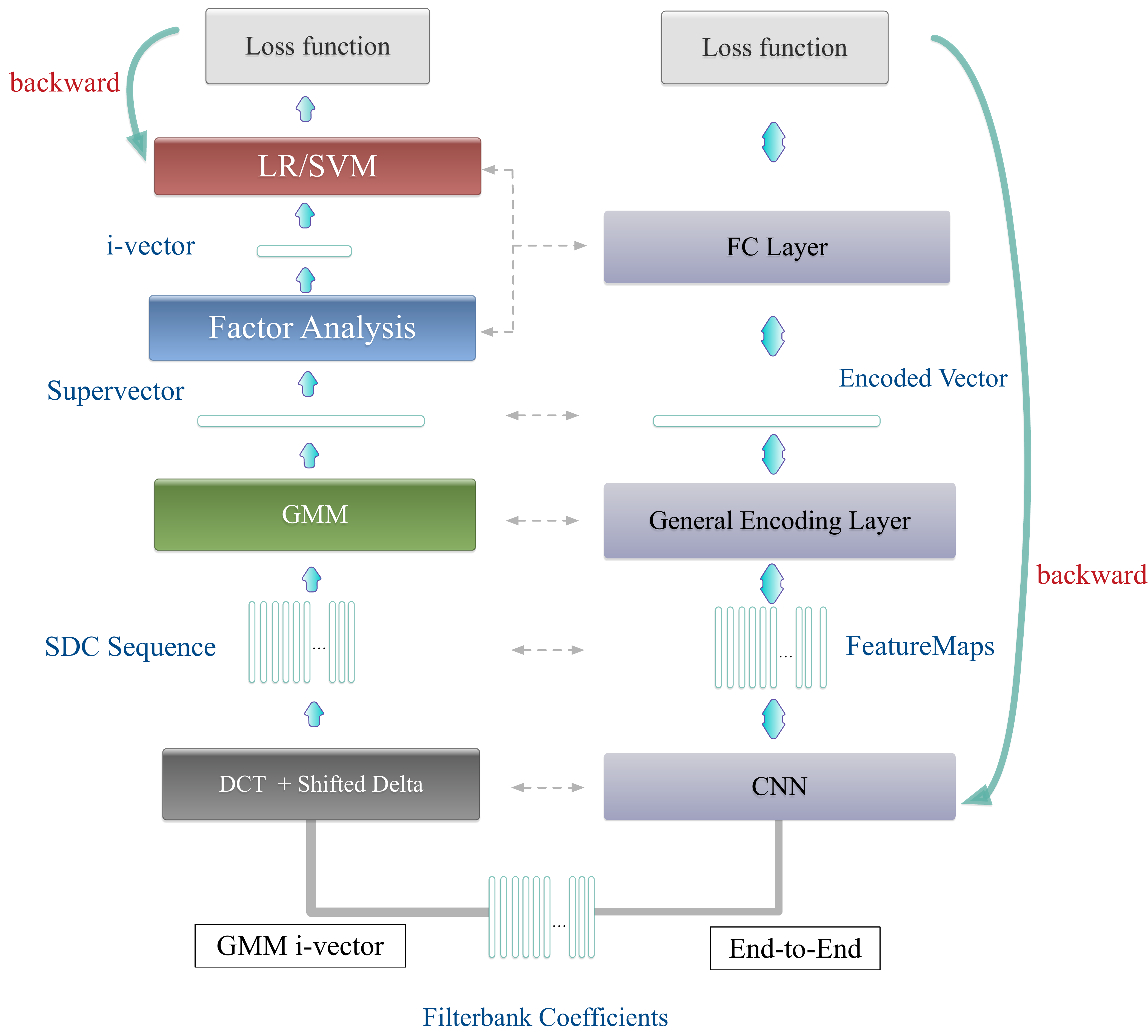}
		\caption{Comparison of classical GMM i-vector approach and the proposed end-to-end learning scheme }\label{fig:comparison}
	\end{figure}

	A comparison of conventional approaches and the proposed end-to-end learning scheme is described in Fig. \ref{fig:comparison}. In traditional methods like GMM i-vector,  each component is optimized in separate steps as illustrated with different colors. In our end-to-end learning scheme, the entire pipeline is learned in an integrated manner because the features, encoding layer and the encoded vector representation for the classifier are all learned jointly. 

	There are two characteristics of our learning scheme showing the theoretical and practical compatibility with the classical GMM i-vector approach: First, each component of our neural network has its parallel equivalent block towards to the classical GMM i-vector processing stream, as demonstrated in Fig. \ref{fig:comparison}.
	Second, our neural network architectures accept variable-length speech inputs and learn an utterance level representation,  which explores the strength of GMM i-vector based methods practically.

	\subsection{The role and effect of CNN}

\begin{figure}[tb]
\begin{minipage}{0.49\linewidth}
 \centerline{\includegraphics[width=4.0cm]{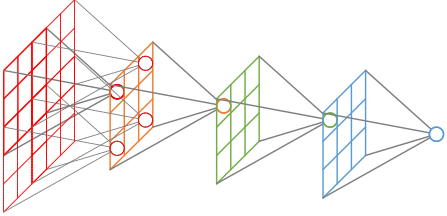}}
\end{minipage}
\hfill
\begin{minipage}{0.49\linewidth}
\centering
\begin{tabular}{|c|c|c|}
  \hline
   layer & shape & RF\\
   \hline
  input & (64, L) & 1 \\
  conv1 & K=(3,3), S=2 & 3\\
  conv2 & K=(3,3), S=2 & 7\\
  conv3 & K=(3,3), S=2 &15\\
  conv4 &K=(3,3), S=2 & 31\\
  conv5 & K=(3,3), S=2 & 63\\
  \hline
\end{tabular}

\end{minipage}
\caption{Example diagram of convolutional receptive field}
\label{toyrecep}
\end{figure}

	CNN is a kind of specialized neural network for processing data with a known grid-like topology and it uses learned filters to convolve the feature maps from the previous layer. One of the basic concepts in deep CNN is the receptive field, as drown in the left part of Fig. \ref{toyrecep}.  A unit in CNN only depends on a region of the input and this region of the input is the receptive field for that unit \cite{Luo2017Understanding}.

	The concept of receptive field is important for understanding how CNN can extract local feature descriptors automatically. Since anywhere outside the receptive field of a unit does not affect the value of that unit, it is necessary to carefully control the receptive field, to ensure that it covers the entire relevant input region. A simple toy CNN is described in right part of  Fig. \ref{toyrecep}.  The input feature coefficient has 64 dimension, with $L$ frames. After five times convolution, the receptive field of output units reach to 63, which not only cover the entire feature coefficients axis, but also a wide temporal context range. 

	To some extent, therefore, the convolution layer of CNNs operates in an sliding window manner acting as a automatic local feature extractor. It learns temporal ordered feature representation automatically under the backward from loss function. 
	
	Any feature sequence with variable-length can fed into CNN theoretically, and in our experiments, Fbank is adopted as the input. 
	
	\subsection{General encoding layer}

	Deep learning is well known for  end-to-end modeling of hierarchical features, so what is the challenge of recognizing language pattern in an end-to-end way and why we need such proposed general encoding layer? 
	
	The output featuremaps of CNN preserve a relative temporal arrangement of the given input feature sequence. A simple  way is to concatenate the resulted globally ordered features together and fed it into the fully-connected (FC) layer for classification. 
	
	However, this is theoretically and practically not ideal for recognizing language, speaker or other paralinguistic information since we need a time-invariant representation instead of concatenation. Therefore, a general encoding layer, which can transfer the temporal ordered frame level local features with variable length into an temporal orderless utterance level vector representation is desirable for our end-to-end learning scheme. 
	
			\begin{figure}[tb]
		\centering
		\includegraphics[width=\columnwidth]{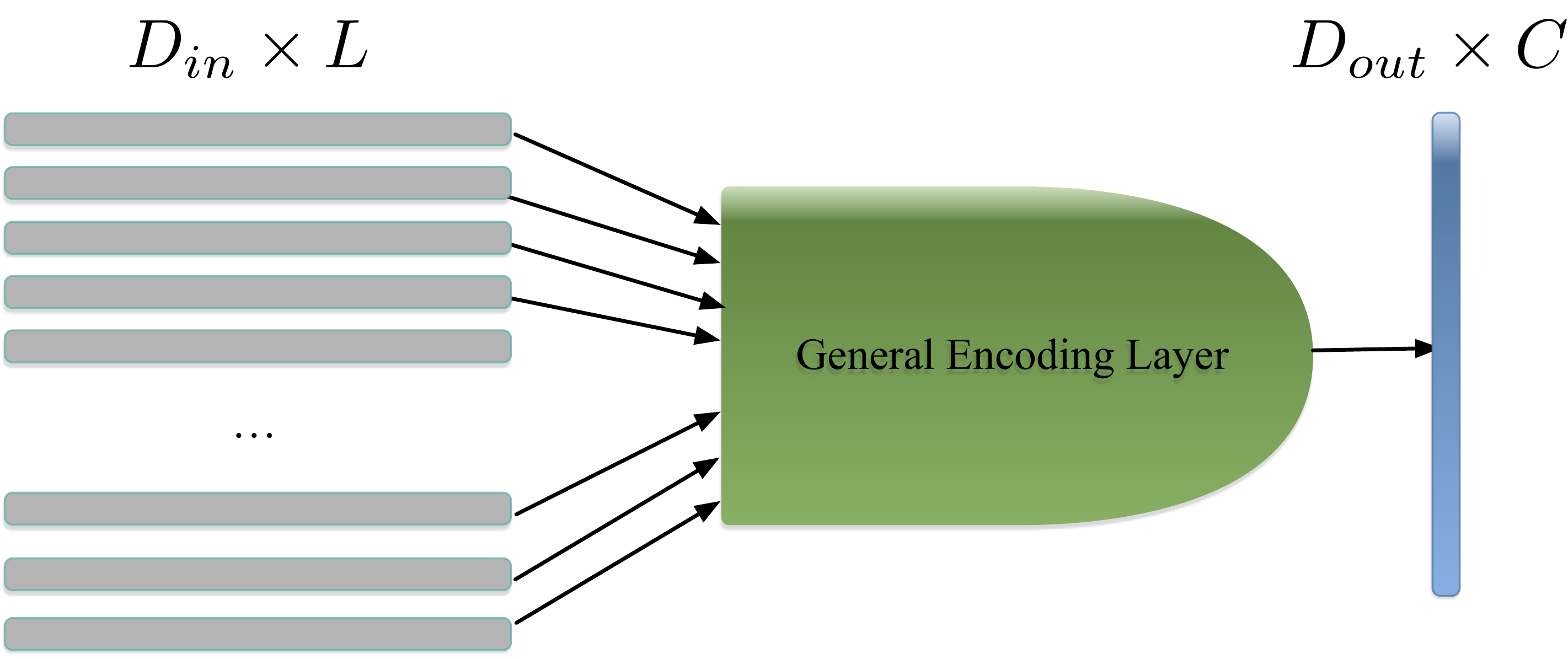}
		\caption{General encoding layer. It receives input feature sequence with variable length, produces an encoded utterance level vector with fixed dimension}\label{fig:encodinglayer}
	\end{figure}

	\begin{figure*}[tb]
	\centering
	\includegraphics[width=0.87\textwidth]{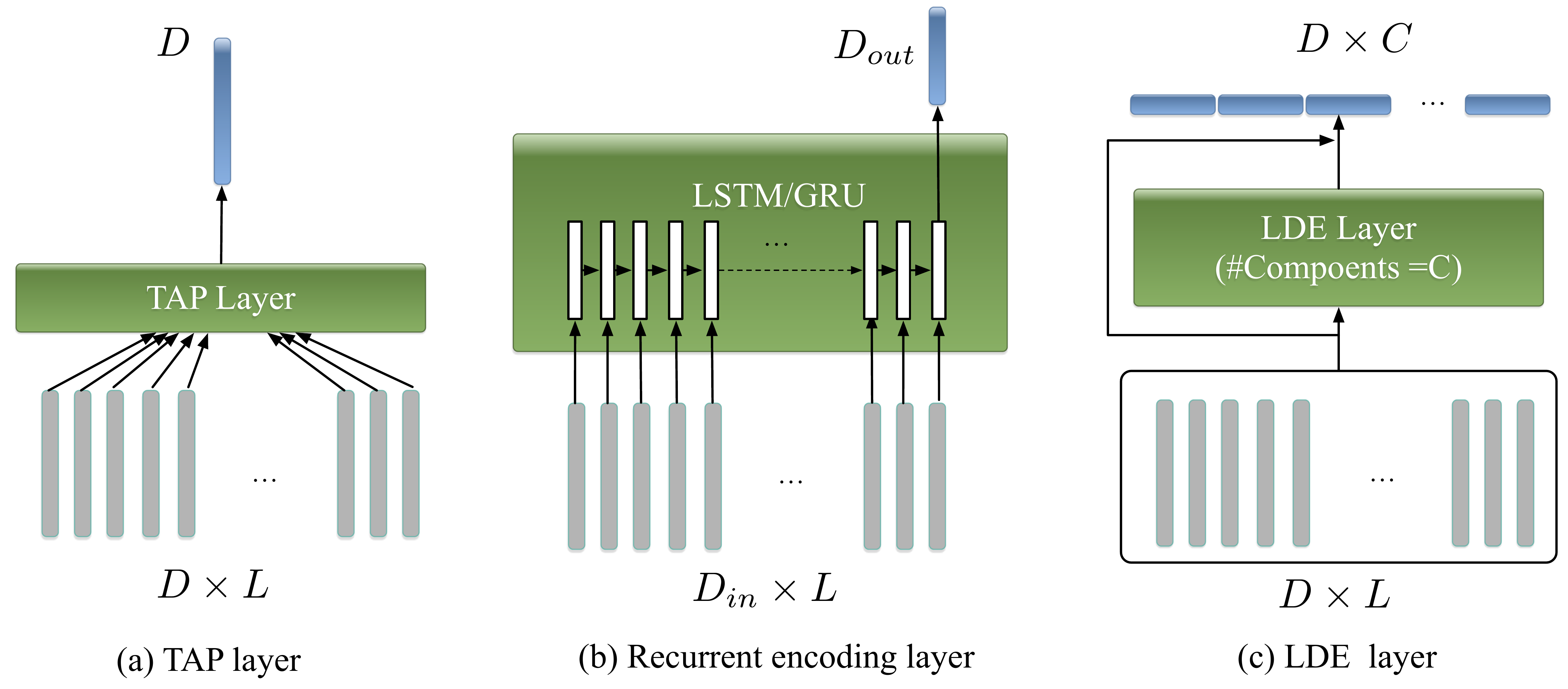}
	\caption{Typical encoding layers. They all receive variable-length sequence, produce encoded utterance level vector with fixed dimension}\label{fig:generalencoding}
\end{figure*}

	The general structure of encoding layer  is demonstrated in Fig. \ref{fig:generalencoding}. For a given input temporal ordered feature sequence with the shape $D_{in} \times L$ (where $D_{in}$ denotes the feature dimension of local feature after CNN, and $L$ denotes the temporal duration length after CNN), the encoding layer aggregates them over time. More specifically, it transforms them into a utterance level temporal orderless $D_{out} \times C$ vector representation, which is independent of duration length $L$.

	We introduce a new axis with shape $C$, because we can imitate the mechanism of conventional K-means/GMM/DNN based dictionary model and design a dictionary to encode vector accumulating high order statistics.

    \section{Architectures}
	\label{sec:architecture}

	\subsection{CNN}
	\label{sec:cnn}
	
	The receptive field size of a unit can be increased by stacking  more layers to make the network deeper or by sub-sampling. Modern deep CNN architectures like Residual Networks \cite{He2016Deep} use a combination of these techniques. Therefore, in order to get higher abstract representation better for utterances with long duration, we design a  CNN based on the well-known deep ResNet  architecture. The detailed network structure is the same as described in  \cite{cailde_iccasp18}.

	\subsection{TAP layer}
	The simplest encoding layer might equally pool the  CNN extracted features over time.  This encoding procedure is context-independent because it just accumulate the mean statistics and doesn't rely on the temporal structure of feature.
	
	\subsection{Recurrent encoding layer}
	The other way to encode vector is to employ recurrent layer , such as long short term Memory (LSTM) layer \cite{Graves2012Long}, gated recurrent unit (GRU) layer \cite{chung2014empirical}. It operates on top of the extracted frame-level features to capture the temporal structure of speech segments into a single representation. This encoding procedure  is context-dependent because it rely on the temporal order of feature sequence.

	Because recurrent layer can  make full use of the context of feature sequences in  forward  directions,  we regard the last output vector of this recurrent layer as the encoded representation and put it only into the following fully-connected layer, as is  in Fig. \ref{fig:generalencoding}.
	
	\subsection{LDE layer}

	The simple average pooling might not be the best way to demonstrate the feature distribution and conventional methods always rely on an dictionary learning procedure like K-means/GMM/DNN, to accumulate more discriminative statistics.

	Inspired by this, we introduce a novel LDE Layer to accumulate statistics on more detailed units. It combines the entire dictionary learning and vector encoding pipeline into a single layer for end-to-end learning.  The LDE Layer imitates the mechanism  of GMM Supervector, but learned directly from the loss function. This representation is orderless which is suitable for text-independent applications, such as LID. The LDE Layer acts as a pooling layer integrated on top of convolutional layers, accepting arbitrary input sizes and providing output as a fixed dimensional  representation. More details can refer to \cite{cailde_iccasp18}.

	\section{Experiments}
	\label{sec:experiments}

	\subsection{Data description}

	For better results reproduction, we conducted experiments on 2007 NIST Language Recognition Evaluation(LRE), because the initial data processing and the referenced  GMM/DNN i-vector baseline systems have  their ready-made recipe developed in the Kaldi toolkit \cite{Povey_ASRU2011}. Our training corpus including  Callfriend datasets,  LRE 2003, LRE 2005, SRE 2008  datasets and development data for LRE07. The total training data is about 37000 utterances. The task of interest is the closed-set language detection. There are totally 14 target languages in testing corpus, which included 7530 utterances split among three nomial durations: 30, 10 and 3 seconds.

	\subsection{Conventional systems}
	As for the  GMM i-vector system, raw audio is converted to 7-1-3-7 based 56 dimensional SDC feature, and a frame-level energy-based VAD selects features corresponding to speech frames. All the utterances are split into  short segments no more than 120 seconds long. A 2048 components full covariance GMM UBM is trained, along with a 600 dimensional i-vector extractor, followed by length normalization and multi-class logistic regression. 
	
	The DNN i-vector system uses an extra DNN phoneme decoder containing 5621 senones, which is trainied with Fisher English database.  The sufficient statistics is accumulated through the DNN senone posteriors and the DNN-initialized UBM. 
	
	To avoid training additional DNN acoustic model, we use DNN PPP feature, rather than DNN bottleneck feature. The 5621 dimensional senone posteriors are converted into  52 dimensional PPP feature  by log transform, principal component analysis (PCA) reduction and a mean-normalization over a sliding window of up to 3 second. For DNN tandem system, the 56 dimensional SDC feature and the 52 dimensional PPP feature is concatenated at feature level to train GMM UBM model. 

	\subsection{End-to-end system}

	Audio is converted to 64-dimensional Fbank with a frame-length of 25 ms, mean-normalized over a sliding window of up to 3 seconds. The same voice activity detection (VAD) processing as in GMM i-vector baseline system is used here.

	The network is trained using a cross entropy loss. The network is trained for 90 epochs using stochastic gradient descent. We start with a learning rate of 0.1 and divide it by 10 and 100 at 60th and 80th epoch. For recurrent GRU/LSTM layer,  a two layer structure with its hidden and output dimension equal to the input vector dimension is adopted. The dictionary component amounts in LDE layer are 64.
	
	For each training step,  an integer $L$ within $\left[ 200 \textrm{,}  1000 \right]$  interval is randomly generated, and each data in the mini-batch is cropped or extended to $L$ frames.

	In testing stage, all the 3s, 10s, and 30s duration data is tested on the same  model. Because the duration length is arbitrary, we feed the testing speech utterance to the trained neural network one by one.

	\subsection{Evaluation}

	Table \ref{table:lre07} shows the performance on the 2007 NIST LRE closed-set task.  
	The performance is reported in  average detection cost $C_{avg}$  and equal error rate (EER).
	
	The conventional system results start from ID1 to ID5, and the DNN PPP feature system achieves the best result.  For ID2 to ID5, additional speech data with transcription and an extra DNN phoneme decoder is required, while our end-to-end systems only rely on the acoustic level feature of LID data. Comparing with GMM i-vector, our end-to-end systems achieve $C_{avg}$ and EER reduction in a large gap. Even comparing with those DNN acoustic model based systems,  our final best CNN-LDE system achieves comparable performance with DNN PPP feature system, outperforming the rest of conventional systems in Table  \ref{table:lre07} significantly.

	It's very interesting that although recurrent layer introduces much more parameters comparing with TAP, it results in a slightly degraded performance.  Specially, when the full 30s duration utterance  is fed into our CNN-GRU/CNN-LSTM neural network trained within 1000 frames (10s), it suffers from ``the curse of sentence length'' \cite{cho2014properties}. The performance drops sharply and almost equals to random guess. In addition, many of previous works \cite{gonzalez2014automatic,Geng2016End} employing  RNN only report  success results on short utterance ($\leq$3s). Therefore, we can infer that although recurrent layer can deal with variable length inputs theoretically, it might be  not suitable for the testing task with wide duration range and particularly with duration that are much longer than those used for training.  
	
	The success of TAP and LDE layer  inspires us that for this kind of paralinguistic speech attribute recognition tasks,  it might be  more necessary to get utterance level representation describing the context-independent feature distribution rather than the temporal structure.
	\begin{table} [tb] 
	\caption{\  Performance on the 2007 NIST LRE closed-set task}
	\centerline{
		\resizebox{0.49\textwidth}{!}{
		\begin{tabular}{c c c c c}
			\hline	
			{ System}& \multirow{2}{*}{ System Description}&\multicolumn{3}{c}{$C_{avg}(\%)/EER(\%)$}\\
			\cline{3-5}
			ID&&3s&10s&30s\\
			\hline
			1&GMM i-vector  &20.46/17.71&8.29/7.00&3.02/2.27\\
			2&DNN i-vector  &14.64/12.04&6.20/3.74&2.601.29\\
			3&DNN PPP  Feature&\textbf{8.00/6.90}&\textbf{2.20/1.43}&\textbf{0.61/0.32}\\
			4&DNN Tandem Feature  &9.85/7.96&3.161.95&0.97/0.51\\
			5&DNN Phonotactic\cite{Gelly2016A}  &18.59/12.79&6.28/4.21&1.34/0.79\\
			\hline
			6&RNN D\&C\cite{Gelly2016A}  &22.67/15.57&9.45/6.81&3.28/3.25\\
			7&LSTM-Attention\cite{Geng2016End}  &-/14.72&-/-&-/-\\
			\hline
			8& \textbf{CNN-TAP}  &9.98/11.28&3.24/5.76&1.73/3.96\\	
			9& \textbf{CNN-GRU} &11.31/10.74&5.49/6.40&-/-\\
			10& \textbf{CNN-LSTM} &10.17/9.80&4.66/4.26&-/-\\
			11& \textbf{CNN-LDE} & \textbf{8.25/7.75}&\textbf{2.61/2.31}&\textbf{1.13/0.96}\\
			\hline
	\end{tabular}}}
	\label{table:lre07}
\end{table}
	\section{Conclusions}
	\label{sec:conclusions}
	In this paper we present a novel end-to-end learning scheme for LID. We give insights into its relationship with classical GMM i-vector in both theory and practice. We analysis the role and effect of each components in the end-to-end pipeline in detail and discuss the reason why they might to be appropriate for LID. The results is 
	very remarkable even though we only use the acoustic level information. In addition,  as is done in conventional i-vector based approach, we can extend the basic learning scheme with phoneme discriminant feature or phonetically-aware dictionary encoding methods in the future. 

	
	\newpage
	\newpage
	\newpage
	\bibliographystyle{IEEEbib}
	\bibliography{csl2}
	
\end{document}